\documentclass[12pt]{article}

\def\cA{{\cal A}}

\newfont{\goth}{eufm10 scaled \magstep1}

\def\a{\alpha}

\def\c{\gamma}\def\C{\Gamma}
\def\d{\delta}
\def\e{\epsilon}

\def\l{\lambda}

\def\th{\theta}

\def\beq{\begin{equation}}\def\eeq{\end{equation}}
\def\beqa{\begin{eqnarray}}\def\eeqa{\end{eqnarray}}
\def\barr{\begin{array}}\def\earr{\end{array}}

\def\del{\partial}

\def\As {{A \hspace{-6.4pt} \slash}\;}
\def\Ds {{D \hspace{-6.4pt} \slash}\;}

\def\ks {{ k \hspace{-6.4pt} \slash}\;}
\def\ps {{p \hspace{-6.4pt} \slash}\;}
\def\pas {{{p_1} \hspace{-6.4pt} \slash}\;}
\def\pbs {{{p_2} \hspace{-6.4pt} \slash}\;}
\def\pt{\tilde{p}}

\let\bm=\bibitem

\def\nn{\nonumber}
\def\bd{\begin{document}}
\def\ed{\end{document}}
\def\ba{\begin{array}}
\def\ea{\end{array}}
\def\bea{\begin{eqnarray}}
\def\eea{\end{eqnarray}}
\def\ft#1#2{{\textstyle{{\scriptstyle #1}\over {\scriptstyle #2}}}}
\def\fft#1#2{{#1 \over #2}}
\newcommand{\be}{\begin{equation}}
\newcommand{\ee}{\end{equation}}
\newcommand{\eq}[1]{(\ref{#1})}
\def\eqs#1#2{(\ref{#1}-\ref{#2})}
\def\det{{\rm det\,}}
\def\tr{{\rm tr}}
\newcommand{\ho}[1]{$\, ^{#1}$}
\newcommand{\hoch}[1]{$\, ^{#1}$}
\def\ra{\rightarrow}
\def\uha{{\hat {\underline{\a}} }}
\def\uhc{{\hat {\underline{\c}} }}

\begin{document}

\hfill{NEIP-00-004}

\hfill{hep-th/0003007}

\vspace{20pt}

\begin{center}

{\Large\bf Induced Chern-Simons 
and WZW action 
in Noncommutative Spacetime }

\vspace{30pt}

{\large Chong-Sun Chu }

\vspace{15pt}

{\small \em Institute of Physics, University of Neuch\^atel, CH-2000
Neuch\^atel, Switzerland}
\vskip .2in \sffamily{chong-sun.chu@iph.unine.ch}

\vspace{60pt}

{\bf Abstract}
\end{center}

We consider noncommutative gauge theory with Dirac or
Majorana fermions in odd
dimensional spacetime and compute the induced 
noncommutative Chern-Simons action generated at 1-loop. 
We observe that 
the Chern-Simons term induced by a  Dirac fermion has a smooth limit 
when $\th \rightarrow 0$, but there is a finite jump for the
Chern-Simons term  induced by a Majorana spinor. 
The induced Chern-Simons action  from a Majorana spinor 
is nonvanishing even in the $\th
\rightarrow 0$ limit, a discontinuity that share a similar 
characteristic as the UV/IR singularity discovered originally 
by Minwalla, Raamsdonk and Seiberg. Properties of the noncommutative
WZW action are also discussed.

\pagebreak 

\section{Introduction }

Chiral fermions in even dimensions can give rise
to an anomaly \cite{anom0}. 
Although there is no chirality in odd dimensions, there
is still a similar phenomenon for fermions. It is
well-known that due to radiative effects at one fermion loop, a
Chern-Simons action for the gauge bosons can be generated in odd
dimensions \cite{fcs1,fcs2}. This phenomenon has wide applications in
particle and condensed matter physics \cite{cond}. 

Recently, noncommutative field theory has shown up as effective
description of string theory in a certain background 
\cite{miao,org,dev,sw}. 
The noncommutativity takes the form
\be
[x^\mu, x^\nu] = i \th^{\mu \nu},
\ee
where $\th^{\mu \nu}$ is a antisymmetric real constant matrix and is
of dimension  length squared.  In the dual language, the algebra of
functions is described by the Moyal product
\be
(f * g) (x) = e^{i \frac{\th^{\mu\nu}}{2} \frac{\del}{\del \xi^\mu}
\frac{\del}{\del \zeta^\nu} } f(x+ \xi) g(x+\zeta) |_{\xi=\zeta=0},
\ee
which is associative,  noncommutative and satisfies
\be
\overline{(f * g)} = \bar{g} * \bar{f}
\ee
under complex conjugation. 
We note also that under integration
\be \label{integ}
\int f*g = \int g*f = \int f g ,
\ee
which is a consequence of momentum conservation. 
Using this $*$-product, 
field theory on a noncommutative spacetime  
can be easily formulated. One simply needs to
replace the usual multiplication of functions by the Moyal product. 
Pioneering analysis of this kind of noncommutative field theory was  
performed by Filk \cite{filk}. Aspects of
noncommutative field theories was further developed in 
\cite{VG}-\cite{anom}. 

In this paper we
analyze gauge theory with fermions defined on a noncommutative 
odd dimensional spacetime
and determine the corresponding induced Chern-Simons action.
Since the action for the noncommutative theory 
is a smooth deformation of the classical action,
one may natively expect to
get back the usual commutative description 
in the limit  $\theta \rightarrow 0$ . 
We find that this is not always the case when one is in the quantum
regime:  for the case of a Majorana spinor, 
there is a jump in the induced Chern-Simons action. 
Singularities in $\th$ have also been
displayed in the scalar theories \cite{seiberg1,irina, seiberg2} and
in QED \cite{ha,mst}.

\section{Induced Chern-Simons in odd dimensions}
 
We will begin with  noncommutative  QED in (2+1)-dimensions
with a 2-components massless  fermion. We will take
$\th_{12} = \th \neq 0$, and  the  action is  given by 
\be
S= \int d^3 x ( -\frac{1}{4} F_{\mu \nu} * F^{\mu \nu} + 
i \bar{\psi} * \Ds *  \psi ),
\ee
where 
$F_{\mu \nu} = \del_\mu A_\nu - \del_\nu A_\mu + i g [A_\mu,
A_\nu]_*$ and the covariant derivative (and hence the fermion
coupling) is given by
$D_\mu \psi = \del_\mu \psi +i g A_\mu * \psi  $ for a Dirac
spinor and $D_\mu \psi = \del_\mu \psi +i g [A_\mu, \psi]_*  $ for a
Majorana spinor. These couplings reproduce the correct corresponding
commutative limit, in particular a Majorana spinor is neutral
in the commutative case. 

The $\c$-matrices are given by the Pauli matrices  
and satisfies 
\be
\c^\mu \c^\nu = -g^{\mu \nu} -i \e^{\mu \nu \l} \c_\l, 
\ee
where $ g^{\mu \nu} = (-,+,+)$ and $\e^{012} =+1$. The action is
invariant under the gauge transformation
\be \label{Atransf}
\d A_\mu = \del_\mu \a - i g [\a,A_\mu]_*, 
\ee
and
$ \d \psi = -i g \a * \psi$ or $\d \psi = -i g [\a , \psi]_* $
for a Dirac or Majorana spinor respectively.

The one fermion-loop effective action is given by
\bea \label{S}
S_{eff} [A,m] =&&  \frac{1}{2} \int \frac{ d^3 p}{(2\pi)^3} A_\mu (p)
A_\nu (-p)  (i \C^{\mu\nu}(p) ) + \nn\\
&&+\frac{1}{3} \int \frac{ d^3 p_1}{(2\pi)^3} \frac{ d^3 p_2}{(2\pi)^3} 
A_\mu(p_1) A_\nu(p_2) A_\l (-p_1-p_2) (i\C^{\mu\nu\l}(p_1,p_2))  
\eea
for $m=0$. The 2-point and 3-point functions $\C_{\mu \nu}(p)$  and
$\C_{\mu \nu\l}(p)$  will be analyzed now.

\subsection{Dirac fermions}

We first consider the case of a Dirac spinor. 
It is crucial \cite{ha}  to observe that for Dirac spinors, 
one only gets planar diagrams from doing  the Wick contractions.
Hence the 2-point and 3-point functions are
\be \label{d2}
i \C_D^{\mu \nu}(p) = g^2 \int \frac{ d^3 k}{(2\pi)^3} 
\tr[ \c^\mu \frac{\ks -\ps -m}{(k-p)^2 +m^2}\c^\nu 
\frac{\ks -m}{k^2 +m^2} ] ,
\ee
\be \label{d3}
i \C_D^{\mu\nu\l}(p_1,p_2)  = - g^3 \int \frac{ d^3 k}{(2\pi)^3} 
\frac{ \tr[ \c^\mu (\ks -m) \c^\nu (\ks+ \pbs -m) \c^\l 
(\ks -\pas -m) ]}
{(k^2 +m^2)((k+p_2)^2 +m^2)((k-p_1)^2+m^2)}  
e^{- \frac{i}{2} p_1 \th  p_2 }.
\ee
The only difference from the commutative case is the phase factor in
\eq{d3} which depends only on the external momenta. 

The effective action \eq{S} is ultraviolet divergent and needs to be
regularized. This can be achieved by the  standard Pauli-Villars
method:
\be
S^{reg}_{eff}[A] = S_{eff}[A,m=0] - \lim_{m\rightarrow \infty} 
S_{eff}[A,m] .
\ee 
As in the commutative case,  
apart from the wanted terms that cure the divergences of \eq{S},
there are terms  
that remain even after sending $m\rightarrow \infty$ and these terms
give rises to the induced Chern-Simons Lagrangian. 
One can verify  that in the large $m$ limit: 
\bea
\lim_{m \rightarrow \infty} i \C^D_{\mu\nu}(p)
&=& -\frac{\Lambda}{3 \pi^2} g_{\mu\nu}
- \frac{i}{4 \pi} \frac{m}{|m|}  \e_{\mu \nu \l} p^\l, \\
\lim_{m \rightarrow \infty} i \C^D_{\mu\nu\l}(p_1,p_2) 
&=& \frac{i}{4\pi}  \frac{m}{|m|} \e_{\mu \nu \l} \;
e^{- \frac{i}{2} p_1 \th  p_2 }.
\eea
Substituting back to \eq{S}, we obtain the following fermion-loop 
induced term in $S^{reg}_{eff}$,
\be \label{ind1}
S_{ind} = \pm \frac{1}{2} S_{CS} 
\ee
with
\be
S_{CS} = \frac{1}{4 \pi} \int d^3 x \; \e^{\mu \nu\l}
(g^2 A_\mu \del_\nu A_\l + \frac{2i}{3}g^3 A_\mu * A_\nu * A_\l ) ,
\ee
or in terms of the gauge field $\cA_\mu = -i g A_\mu$
\be \label{CS}
S_{CS} = \frac{1}{4 \pi} \int d^3 x \; \e^{\mu \nu\l}
( \cA_\mu \del_\nu \cA_\l + \frac{2}{3} \cA_\mu * \cA_\nu * \cA_\l ) 
\ee
is the noncommutative Chern-Simons action. $S_{CS}$ is local invariant
under \eq{Atransf}, while under a finite gauge transformation,
\be \label{Atransf1}
\cA_\mu \rightarrow h^{-1}_* * \cA_\mu *h + h^{-1}_* * \del_\mu h
\ee
$S_{CS}$ changes as
\footnote{ We have dropped a total derivative piece $ d(A * dh*
h_*^{-1})$ which on a manifold with boundary will give rises to an
anomaly for the boundary theory. Descent relations and
the algebraic structures of chiral anomaly still hold generally for a
noncommutative gauge theory \cite{zumino}. See also the third
and fourth references of \cite{anom} for recent discussion.
}
\be \label{cc1}
S_{CS} \rightarrow S_{CS}  - 2 \pi w
\ee
where 
\be
S_{WZW} := \frac{1}{24 \pi^2} \int_B d^3 x \;  \e^{\mu \nu \l}
(h^{-1}_* *\del_\mu h * h^{-1}_* *\del_\nu h * h^{-1}_* * \del_\l h)
\ee
is the noncommutative WZ term over $B$ 
(or WZW action over $\del B$) and $w:=S_{WZW}$ for $B=S^3$ is the
``winding number''.  
Here $h^{-1}_*$
is the inverse of $h$ with respect to the Moyal product:
\be
h *  h^{-1}_* = h^{-1}_* * h =1 .
\ee
For any $h =e^{i \a}$ of  $U(1)$, it is  $h^{-1}_* =e^{-i \a} $. 
It is easy to check that $w$ is  invariant under an infinitesimal
transformation $\d h = \l * h$
and that under a finite transformation,
\be\label{finitetransf}
w(h'*h) = w(h') + w(h).
\ee
Now we claim that $w$ is zero for the Abelian case.
Consider 
\be
I[\a] = \e^{\mu \nu \l} \int d^3 x \;   e^{-i \a} * \del_\mu e^{i\a} 
* e^{-i \a} * \del_\nu e^{i\a} * e^{-i \a} * \del_\l e^{i\a}. 
\ee
It is clear from \eq{finitetransf} that 
$I[\frac{m}{n} \a] =\frac{m}{n}I[\a]$ for any integers $m,n$, therefore
$ I[s \a] = s I[\a]$ for any real constant $s$,
and hence 
\be \label{e1}
I[\a] = \frac{\del}{\del s} I[s \a] .
\ee
But we already knew that $I$ is invariant under
arbitrary infinitesimal transformation, hence our claim.  One can also
get the same result by noticing that since 
the LHS of \eq{e1} is independent of $s$, one can 
evaluates the RHS of \eq{e1} at the particular value $s=0$ and obtains
the desired result.  That $w=0$ for $U(1)$ may  be expected
intuitively since  $S^3$ is too big to fit in $S^1$. As a result, the
2-dimensional noncommutative WZW action \cite{moreno} is well defined.

For the case of non-Abelian $U(N)$ gauge fields, we just have to
replace \eq{S} by 
\bea \label{S_nonabelian}
&& S_{eff} [A,m] =  \frac{1}{2} \tr(T^a T^b) 
\int \frac{ d^3 p}{(2\pi)^3} A^a_\mu (p)
A^b_\nu (-p)  (-i \C^{\mu\nu}(p) ) \nn\\
&&+\frac{1}{3} \tr(T^a T^b T^c) 
\int \frac{ d^3 p_1}{(2\pi)^3} \frac{ d^3 p_2}{(2\pi)^3} 
A^a_\mu(p_1) A^b_\nu(p_2) A^c_\l (-p_1-p_2) (-i\C^{\mu\nu\l}(p_1,p_2))  
\eea
with the same 2-point and 3-point functions. Thus one again obtains 
\eq{ind1}, now with 
\be \label{nonabCS}
S_{CS} = \frac{1}{4 \pi} \int d^3 x \; \e^{\mu \nu\l}
\tr (\cA_\mu \del_\nu \cA_\l + \frac{2}{3} \cA_\mu * \cA_\nu * \cA_\l ),
\ee
where $ \cA_\mu = -i g A_\mu^a T^a$.

Again, $S_{CS}$ is local gauge invariant, while 
under a finite gauge transformation \eq{Atransf1} with $h$ in $U(N)$,
$S_{CS}$ changes as
\be
 S_{CS} \rightarrow S_{CS}  - 2 \pi w
\ee
where 
\be
S_{WZW} := \frac{1}{24 \pi^2} \e^{\mu \nu \l}\int_B \tr 
(h^{-1}_* *\del_\mu h * h^{-1}_* *\del_\nu h * h^{-1}_* * \del_\l h)
\ee
is the noncommutative WZ term over $B$ and   $w:=S_{WZW}$ for $B=S^3$.
It is easy to check that 
$w$ is  invariant under a local $U(N)$ gauge
transformation $\d h = \l* h $ and we have again \eq{finitetransf}
under a finite transformation.
It is not clear whether $w$ is an integer or not when integrated
on $S^3$. One can shows that this number is independent of $\th$, 
at least to the first order in $\th$. 
Since $w[h^{*n}] = n w[h] $ and $w$ is invariant
under small changes of the map $h$, these suggest that $w$ may
again serve as some sort of homotopy invariant and that the
$\th$-dependence factorize.
If $\th$ dependence does not 
disappear and $w$ is not an integer, then
one will need to arrange the fermions content so that the global anomaly
cancel. 
We leave these interesting issues for future investigation. 

\subsection{Majorana fermions}

We next consider the case of a Majorana spinor. The situation differs
in that now nonplanar diagram can also contribute. 
Similar considerations have also been made in \cite{mst}.  
The Feynman rules have been worked out in \cite{ha} 
and we will not repeat them here. 
We note that the phase factor  $e^{-i q_1 \frac{\theta}{2} q_2}$   
for a Dirac spinor coupled to the gauge field (with momentum $q_1,
q_2,q_3$ coming into the vertex) is now replaced by 
$-2i \sin(q_1 \frac{\theta}{2} q_2)$ due to 
the commutator nature of  the coupling for the  Majorana spinors.
In particular the 2-point and 3-point functions are
now given by 
\be \label{m2}
i \C_M^{\mu \nu}(p) = -4g^2 \int \frac{ d^3 k}{(2\pi)^3} 
\tr[ \c^\mu \frac{\ks -\ps -m}{(k-p)^2 +m^2}\c^\nu 
\frac{\ks -m}{k^2 +m^2} ] \sin^2(\frac{\pt k}{2}) ,
\ee
\bea \label{m3}
i \C_M^{\mu\nu\l}(p_1,p_2)  = -8 i g^3 \int \frac{ d^3 k}{(2\pi)^3} 
\frac{ \tr[ \c^\mu (\ks -m) \c^\nu (\ks+ \pbs -m) \c^\l 
(\ks -\pas -m) ]}
{(k^2 +m^2)((k+p_2)^2 +m^2)((k-p_1)^2+m^2)}  \nn\\
\cdot \sin(\frac{\pt_1 k}{2}) \sin(\frac{\pt_2 k}{2}) 
\sin(\frac{\pt_3 (k+ p_2)}{2}) ,
\eea
where for convenience we have denoted $\pt = p \th $ and momentum
conservation 
\be \label{momcons}
p_1+p_2+p_3=0 
\ee 
has to be used. Since 
\be \label{factor1}
\sin^2(\frac{\pt k}{2}) = \frac{1}{2}( 1- \cos \pt k ),
\ee
one can reduce \eq{m2} to a sum corresponding to planar and nonplanar
contributions 
\be
\C_M^{\mu\nu} = -2 \C_D^{\mu\nu} + \mbox{``nonplanar''},
\ee
where ``nonplanar'' are expressions of the form 
\be
 \int \frac{ d^3 k}{(2\pi)^3}
\frac{f(k)}{ ((k-p)^2 +m^2) (k^2 +m^2) } e^{i k \pt }
\ee
and $f(k) = k^\mu k^\nu$, $k^\mu$ or 1. It is easy to show that the
``nonplanar'' terms all vanish in the large $m$ limit. 
It is enough to calculate for the case of $f=1$ as the others can
be obtained by differentiating with respect to $\pt$.
Introducing Schwinger parameters, one obtains
\bea
\int \frac{ d^3 k}{(2\pi)^3}
\frac{1}{ ((k-p)^2 +m^2) (k^2 +m^2) } e^{i k \pt } & =&
\frac{1}{(2 \sqrt{\pi})^3}
\int_0^\infty dT \frac{1}{T^{1/2}} e^{ -m^2 T -\frac{\pt^2}{4T}}
\int_0^1 dx e^{-T x(1-x) p^2}  \nn\\
& < &  \int_0^\infty dT \frac{1}{T^{1/2}} e^{ -m^2 T
-\frac{\pt^2}{4T}} 
\quad \sim \frac{1}{m} e^{-|m| |\pt|}
\eea
and hence
\be
\lim_{m \rightarrow \infty} \mbox{``nonplanar''} =0
\ee
as long as $|\pt| = |\th| \sqrt{p_1^2+p_2^2} \neq 0$, which is of
measure zero in the integration of \eq{S}.
Similarly one gets using  \eq{momcons}
\be \label{factor2}
i \sin(\frac{\pt_1 k}{2}) \sin(\frac{\pt_2 k}{2}) 
\sin(\frac{\pt_3 (k+ p_2)}{2}) = \frac{i}{4} 
\sin (\frac{p_1\th p_2}{2})   
+ \mbox{phases involving internal momenta}.
\ee
Therefore again we can write \eq{m3} as
\be
\C_M^{\mu\nu\l} = -2 \C_D^{\mu\nu\l} +  ``\mbox{nonplanar}''
\ee
and similarly show that the ``nonplanar'' terms vanish in the large
$m$ limit. Substituting back to \eq{S} and remember that there is now
an extra factor of $1/2$ since  we are considering Majorana spinor and
hence $S_{eff} = 1/2 \tr \log (i \Ds + \As)$.
We obtain finally the induced term
\be \label{ind2}
S_{ind} = \pm \frac{1}{2} S_{CS}.
\ee
We note that the factor of 1/2 and 1/4 in \eq{factor1} and
\eq{factor2} can be easily understood. There are one planar and one
nonplanar diagram contributing to 
the 2-point function, and one planar and three
nonplanar diagrams   contributing to the 3-point function (given a
definite cyclic order of the external momentum). 
We also note that in the commutative case, Majorana spinors are 
neutral and
there is no Chern-Simons action induced from them. 
In the noncommutative case, a coupling can be written down 
that reduces to zero in the commutative limit. Both planar and
nonplanar diagrams contribute to the effective actions, but 
the nonplanar diagrams are suppressed and
only the planar diagrams survive the large $m$ limit. Therefore upon
removing the Pauli-Villars regulator, we are left with the induced
Chern-Simons action \eq{ind2}. 


For the $U(N)$ case, the coupling is replaced by
\be
-ig ( f^{abc} \cos (q_1 \frac{\th}{2} q_2) + 
d^{abc} \sin (q_1 \frac{\th}{2} q_2) )
\ee
where we have adopted the normalization $\tr T^a T^b = \d^{ab}/2$ and
$T^a T^b = \frac{1}{2}(i f^{abc} T^c + d^{abc} T^c)$. 
The $U(1)$ generator is $T^0 = \frac{1}{\sqrt{2N}} 1$ and it is 
$f^{000}=0$ and $d^{000}=\sqrt{2/N}$. 
The previous $U(1)$ case is recovered by 
$g d^{000} \rightarrow 2g $. 
Using the identities
\bea
&- f^{a b' a'} f^{b c' b'} f^{c a' c'} = 
f^{a b' a'} d^{b c' b'} d^{c a' c'} =
\frac{N}{2} f^{abc}, \\
&d^{a b' a'} d^{b c' b'} d^{c a' c'} = \frac{N}{2} d^{abc}
+ \d^{ab} \tr T^c +\d^{bc} \tr T^a +\d^{ca} \tr T^b 
\\
&f^{a b' a'} f^{b c' b'} d^{c a' c'} = -\frac{N}{2} d^{abc}
- \d^{ab} \tr T^c +\d^{bc} \tr T^a +\d^{ca} \tr T^b 
\eea
where $T^a$ are in the defining representation. One easily obtains the
induced Chern-Simons term \eq{ind2}, with now $S_{CS}$ 
\eq{nonabCS} defined in the adjoint representation. 

Finally we remark that 
all of the above can be generalized  straightforwardly to
higher dimensions (provided that massive Majorana spinors exist in that
dimensions ) 
and one still obtains the same shift
with the corresponding higher dimensional  Chern-Simons action.

\section{Discussions}

In this paper, we investigated the induced Chern-Simons action 
due to Dirac
and Majorana fermion coupled to Abelian $U(1)$ or non-Abelian  
$U(N)$ gauge fields. The surprising result is that 
for the Majorana spinor case, the induced term does not go to
zero as $\th \rightarrow 0$ and displays a discontinuity. 
Since there is a finite difference no
matter how small is the noncommutativity, this kind of discontinuity 
in a physical quantity may be  useful to searching of 
experimental signals  of noncommutativity and 
Lorentz symmetry breaking in nature. 

Similar $\th$-singularities have also been discovered in scalar field
theories \cite{seiberg1,irina, seiberg2} and QED \cite{ha,mst}.  
These examples show that although classically noncommutative 
physics is a smooth deformation of the commutative description, 
there can be important new quantum mechanical terms
that do not vanish in the commutative limit. The commutative limit and
the classical limit generally do not commute. 
The technical reason is very simple, generally 
one cannot exchange the order of taking the $\th \rightarrow 0$ 
limit and the integration. 

One may then get the opposite impression that 
noncommutative quantum physics always do not
have a smooth commutative limit. The example of the induced 
Chern-Simons term
for Dirac spinors  shows that this is  not true. So long as only
planar diagrams contribute, there is a smooth limit. 
Take another example, the computation of the chiral anomaly for Weyl
fermion. Again only planar diagrams contribute and so the loop
effects get  modified only by a phase factor which depend on the
external momentum. Therefore it can be Fourier transformed  back
easily and gives the usual expression of the anomaly, 
except that now the
products are replaced by the Moyal products.  There is again a smooth
limit because only planar diagrams contribute. 

The opposite situation would be more interesting: is there a case where
only nonplanar diagrams contribute? From the string point of view, we
need a string theory with only nonplanar worldsheets. 
One needs to identify a limit so
that only nonplanar worldsheets survive. Consistency may be an
important issue, but it may also be possible that 
the field theory limit is
not sensitive to the details of the string consistency issues.

It is known that in the commutative case, the Chern-Simons action 
can also be induced from the gauge boson loops. This bosonic 
shift  has been shown
explicitly in the $(2+1)$-dimensional case by Witten \cite{wit} using a
saddle point analysis. The situation is however less clear if one
perform a perturbative calculation and the shift depends 
on the regularization scheme. A natural regularization
is to use string theory. 
It would be interesting \cite{csc2} to see what
string regularization has to say about the shift for both the commutative
and noncommutative case. 
 
Another interesting question concerns the nonrenormalization  of our
result. In the commutative case, 
it has been shown through explicit calculations 
that the fermion induced Chern-Simons term 
is not renormalized at two loops \cite{yc}, for both the Abelian and
non-Abelian case. Beyond this result, we have the Coleman-Hill theorem
\cite{coleman} 
for the Abelian case which shows that there is no contribution to the
induced Chern-Simons action beyond one loop. Due to the 
topological origin of the non-Abelian Chern-Simons action and its
connection with the chiral anomaly, it is also  
expected that it should not be renormalized beyond one-loop. Thus
the situation is more or less 
clear. It would be interesting to investigate
the status of  nonrenormalization theorem for induced Chern-Simons
action, as well as  for chiral anomaly.

Induced Chern-Simons term in odd dimensions 
were originally \cite{fcs1,fcs2} obtained
for fermions coupled to gauge fields in a flat spacetime. This has been
extended \cite{AGPM} 
to full generality for arbitrary curved backgrounds and any
odd dimensions and is shown to be related to the Atiyah-Patodi-Singer
index theorem. The induced Chern-Simons action in $(2n+1)$-dimensions is
given (up to a normalization factor)
by the secondary characteristic class
$Q(A,\omega)$ satisfying
\be \label{Q}
dQ(A,\omega) = {\hat A(R)} ch(F) |_{2n+2},
\ee
where $\omega$ is the gravitational connection. In view of  the 
results \cite{anom} on
the mathematical structure of anomalies in
noncommutative gauge theory, it seem appropriate 
to investigate the deformation theory 
for elliptic operators due to a Moyal product and to study the
possible ``topological'' meaning of the Chern-Simons action and chiral
anomaly and to establish the corresponding index theorems. Better
understanding of the properties of the noncommutative WZW action and
its topological meaning is
also highly desired.

It would also be interesting to investigate aspects of AdS/CFT
correspondence \cite{ads,adsanom} in relation to chiral anomaly and
Chern-Simons action. This will provide another channel to understand
better this so far rather poorly understood correspondence. 
We plan to come back to these issues in the future. 

\bigskip
\vspace{.5cm}

\noindent{\large \bf Acknowledgments}

I would like to thank A. Bilal, J.-P. Derendinger, Rodolfo Russo 
and Bruno Zumino
for many useful discussions and comments. 
This work was partially supported by the Swiss National Science
Foundation, by the European Union under TMR contract
ERBFMRX-CT96-0045 and by the Swiss Office for Education and
Science.


\ed